\documentclass[aps,prd,twocolumn,showpacs,superscriptaddress,floatfix,10pt]{revtex4}
\usepackage{natbib}
\usepackage{amssymb}
\usepackage{graphicx}
\expandafter\let\csname equation*\endcsname\relax
\expandafter\let\csname endequation*\endcsname\relax
\usepackage{amsmath}
\usepackage{amsbsy}
\usepackage{amsthm}
\usepackage{bbm}
\usepackage{bm}
\usepackage{epsfig}
\usepackage{natbib}
\usepackage{dsfont}
\usepackage{hyperref}
\usepackage{soul}
\usepackage[usenames,dvipsnames]{xcolor}

\begin{document}

\title{Spacetime effects on satellite-based quantum communications}
\date{\today}

\author{David Edward Bruschi}\thanks{Current affiliation: Racah Institute of Physics and Quantum Information Science Centre, the Hebrew University of Jerusalem, Givat Ram, 91904 Jerusalem, Israel}
\affiliation{School of Electronic and Electrical Engineering, University of Leeds, Leeds LS2 9JT, United Kingdom}

\author{Timothy C. Ralph}
\affiliation{School of Mathematics and Physics, University of Queensland, Brisbane, Queensland 4072, Australia}

\author{Ivette Fuentes}
\affiliation{School of Mathematical Sciences, University of Nottingham, University Park, Nottingham NG7 2RD, United Kingdom}

\author{Thomas Jennewein}
\affiliation{Institute of Quantum Computing and Department of Physics and Astronomy, University of Waterloo, 200 University Avenue West, Waterloo N2L 3G1, Ontario, Canada}

\author{Mohsen Razavi}
\affiliation{School of Electronic and Electrical Engineering, University of Leeds, Leeds LS2 9JT, United Kingdom}

\begin{abstract}
We investigate the consequences of space-time being curved on space-based quantum communication protocols. We analyze tasks that require either the exchange of single photons in a certain entanglement distribution protocol or beams of light in a continuous-variable quantum key distribution scheme. We find that gravity affects the propagation of photons, therefore adding additional noise to the channel for the transmission of information. The effects could be measured with current technology.
 \end{abstract}

\maketitle

\section{Introduction}
The past century can be regarded as the beginning of the information era. Information science has played a key role in many fields of science, in the development of new technologies and within almost every other human endeavor. At its core, information science aims at understanding how to efficiently encode, transmit, store, manipulate and retrieve information \cite{Shannon:48}. Although great progress in this field was made considering classical physics alone, in the last decades it was shown that quantum mechanics can bring the game to the next level \cite{Scarani:BechmannPasquinucci:09,Alleaume:Bouda:07}. With the development of quantum information science, a new wealth of protocols and devices have been proposed and it has been shown that fundamental limits valid within the realm of classical physics can now be surpassed \cite{Bennett:Brassard:84,Shor:94}. Quantum information science has now been developed up to the point where commercial applications are available. For example, quantum key distribution (QKD) aims at sharing a secret key among two legitimate users, which can be used to achieve secure transfer of information. In this work we will analyze quantum communication protocols where the consequences of space-time being curved play an important role.

With the turn of the century, several space agencies have shown interest in developing and implementing quantum communication networks based on technologies such as quantum relays \cite{Duan:Lukin:01}. Several proposals for quantum communications within low earth orbits (LEOs) have been made, such as the SPACEQUEST and QEYSSAt projects \cite{Armengol:Furch:08,Ursin:Jennewein:09,Rideout:Jennewein:12}.
Most proposed systems have been studied using quantum optics, with little attention  given to the theory of relativity, which describers phenomena that occur at large scales in the presence of gravitational fields \cite{Rideout:Jennewein:12,Ursin:Jennewein:09,Zych:Costa:11}.
It is therefore of practical as well as fundamental importance \cite{Rideout:Jennewein:12} to study such effects on communication protocols when the parties involved (e.g., satellites) are located at great distances within curved space times.

The novel field of relativistic quantum information aims at understanding how relativity affects quantum information tasks \cite{Ralph:Downes:12,Alsing:Fuentes:12}. Most of hitherto research has focused on modeling and employing localized systems for quantum information processing \cite{Bruschi:Fuentes:12,Friis:Lee:13,Friis:Huber:12}, while only recently some attention has been drawn towards understanding the influence of gravity on quantum protocols. For example, the recent work in \cite{Friis:Lee:13} shows that when two users employ the modes of quantum fields contained within cavities to perform a teleportation protocol, the motion of one cavity affects the final fidelity of teleportation. In \cite{Zych:Costa:12} it was shown how curvature could effect a large scale photonic interferometer.

%The first attempt to understand the effects of relativity on simple communication protocols (in particular continuous-variable QKD protocols), was undertaken in \cite{Downes:Ralph:13}. There, two observers in flat space-time tried to communicate by exchanging localized beams of light, i.e., coherent states built of wave packets of light, while one observer was inertial and the other was uniformly accelerated. It was shown that due to different observers disagreeing on the notion of particle, encoded in the well known Bogoliubov transformations \cite{Birrell:Davies}, communication is affected more with increasing acceleration. A simple numerical analysis concluded that accelerations greater than $10^{21}$g were needed to measure any effect \cite{Downes:Ralph:13}.
%It, however, remains too hard with current technology to achieve such regimes of accelerations where such effects become visible.

In this work we show how gravity affects quantum communication protocols and that the effects can be measured with current technology.
In particular, we are interested in the effects of the Earth's gravitational field on quantum communications between ground and space links.
The main framework that naturally allows to investigate phenomena lying at the intersection of quantum mechanics and relativity is quantum field theory (QFT) \cite{Srednicki}. We develop the necessary techniques that allow us to revisit communication protocols between users Alice and Bob located at different heights in a (non uniform) static gravitational potential. In particular, we investigate photon propagation in Schwarzschild space-time, which well approximates the space-time outside non rotating spherical planets. We will use quantum optical models for communication generalized to the relativistic QFT scenario like those developed in \cite{Downes:Ralph:13}. However, contrary to the work in \cite{Downes:Ralph:13}, where the main result depends on two observers disagreeing on the notion of particle (mathematically implemented by non-trivial Bogoliubov transformations), in our work, different observers agree on the particle content of a quantum state. For example, a photon created by Alice reaches Bob intact as a single photon. However, we will study how single-photon wave packets exchanged by users are affected by the gravitational potential, and how this effect impacts protocols such as entanglement distribution, swapping, or QKD \cite{Duan:Lukin:01}. We show that an Earth-to-space QKD system that relies on entanglement distribution using photons could have an additional contribution to its quantum bit error rate (QBER) of as high as $0.7\%$ as a consequence of space-time curvature. This effect would be observable with current technologies.

We suggest that it is possible to correct for the effects of gravity by employing extra resources. For example, we show that Alice and Bob might use an extra reference beam (a local oscillator) during their communication with Gaussian states. Such extra resources cannot be local but Alice and Bob need to exchange extra information in order to apply the desired corrections. This can, in principle, have a substantial impact on the complexity and performance of any quantum communication protocols.

This work is organized as follows. In section \ref{tools} we introduce the tools necessary to address the effects of gravity on quantum communication protocols. In section \ref{photons} we model the creation, propagation on curved background and reception of single photons. In section \ref{photons:traveling} we quantify the effects of curvature on propagating photons. In section \ref{protocol} we apply the tools developed in the previous sections to analyze two different quantum communication protocols. Finally, in section \ref{results} we quantify the effects for realistic configurations. Throughout the paper we take the conventions $\hbar=c=1$. We use the Einstein summation convention i.e., contracted indices are summed over.

\section{Background tools\label{tools}}
In order to study the effects of gravity on quantum communication protocols we will employ tools from QFT and general relativity. On the one hand, QFT provides the description of quantum systems that propagate on a curved but otherwise classical background space-time. On the other hand, general relativity describes the background space-time itself. This standard, but largely experimentally untested approach is referred to as QFT on a curved background \cite{Birrell:Davies}. In this section we will provide a the model for the space-time outside a non-rotating planet and the single photons that propagate from the surface to outer space.

\subsection{Space-time outside a planet}
The space-time outside a spherical non-rotating body can be modeled by $(3+1)$-dimensional Schwarzschild space-time \cite{Misner:Thorne:73,Wald:84}. Standard Schwarzschild coordinates $x^{\mu}$ are $(t,r,\theta,\phi)$ where $t,r$ represent the proper time and radius for observers that are (infinitely) far from the origin $r=0$ (also known as the singularity) \cite{Misner:Thorne:73,Wald:84}. The spherical planet responsible for the curvature has mass $M$, vanishing angular momentum $J$ and radius $r_E$. The Schwarschild metric $g_{\mu\nu}$ in the vacuum outside the planet is
\begin{align}
g_{\mu\nu}=\text{diag}\left(-f(r),\frac{1}{f(r)},r^2,r^2\sin\theta\right)\label{schwarschild:metric}
\end{align}
where $f(r):=1-\frac{r_s}{r}$, $r_S:=\frac{2GM}{c^2}$ is the Schwarzschild radius for the planet and $G$ is the gravitational constant. We assume that the planet's radius $r_E$ is much larger than its Schwarzschild radius $r_S$, i.e. $r_E\gg r_S$. This is the case, for example, for the Earth where $r_S/ r_E\sim1.4\times10^{-9}$.
Inside the planet, the metric depends on the particular model that describes the planet's matter and its distribution \cite{Misner:Thorne:73}. Since we consider communication outside the planet we are not interested in the space-time for $r<r_E$.

The main effects of gravity for the scenarios of interest in this work will depend on the Schwarschild radius $r$. It is well known that field equations with the metric \eqref{schwarschild:metric} can be solved by separating the solution into temporal, radial and angular part. The full solution to our problem requires working with $3+1$ dimensions and the differential equation for the radial part yield no analytical solution \cite{Wald:84}. In contrast, $1+1$ dimensional Schwarzschild contains all of the essential physical properties of its $3+1$ counterpart, while allowing for simple and analytical formulas. We can therefore assume that the problem is essentially $1+1$ dimensional and that the angular part does not contribute to the effects of interest. This will be reasonable provided we only consider radial communication and detector and sources are assumed small compared to $r.$The Schwarzschild metric $g_{\mu\nu}$ in $1+1$ dimensions reads
\begin{align}
g_{\mu\nu}=\text{diag}\left(-f(r),\frac{1}{f(r)}\right)\label{reduced:schwarschild:metric}.
\end{align}
and the line element $ds^2$ in Schwarzschild coordinates is
\begin{align}
ds^2:=g_{\mu\nu}dx^{\mu}dx^{\nu}=-f(r)dt^2+\frac{1}{f(r)}dx^2\label{line:element}.
\end{align}
General relativity predicts that, in the absence of forces, test particles follow geodesics \cite{Misner:Thorne:73}; in $1+1$ Schwarzschild they coincide with test particles free falling towards the origin $r=0$. In order to stay at some constant distance from the planet (i.e. $r=$const) an observer needs to employ some source of acceleration (i.e. a rocket) which counters the gravitational potential. Such trajectory a $r=$const is not a geodesic. In $3+1$ Schwarzschild new geodesics of constant Schwarzschild radius exist, i.e. circular orbits. In this case no acceleration is needed but the observer must have angular momentum. Satellites for standard communication or the global positioning system (GPS) typically follow such orbits. Furthermore, experiments in LEO orbits have recently been proposed to test the effects of gravity on entanglement \cite{Bruschi:Sabin:13}. In this case, the effects of special relativity, specifically the relative motion of two parties, might contribute to the final effect.

An observer at constant distance $r_0$ from the (center of the) planet employs his or her own clock to measure the time in his or her rest frame. The proper time $\tau$ is related to the Schwarzschild time coordinate $t$ by
\begin{eqnarray}
d\tau^2&:=&\left.\frac{ds^2}{c^2}\right|_{r=r_0}=-f(r_0)dt^2,
\end{eqnarray}
which can be simply integrated to give
\begin{eqnarray}
\tau=\sqrt{f(r_0)}t.\label{proper:time}
\end{eqnarray}
Equation \eqref{proper:time} gives the relation between the proper time $\tau$ of an observer sitting at a constant coordinate $r_0$ and the proper time $t$ of an observer (infinitely) far from the planet.

\subsection{Modeling Quantum Optics}
In order to study communication protocols that employ the exchange of pulses of light at the quantum level, the optical pulses can be modelled by wave packets built of monochromatic modes (i.e., plane waves) of an uncharged massless scalar field operator $\Phi(t,x)$ \cite{Srednicki}. It is well known that uncharged scalar fields are a good approximation to the longitudinal (or transverse) modes of the electromagnetic field \cite{Srednicki,Friis:Lee:13}. The field $\Phi$ obeys the standard massless Klein-Gordon equation
\begin{eqnarray}
\square\Phi=0\label{klein:gordon:equation},
\end{eqnarray}
where the d'Alambertian $\square$ in curved space times is defined as
$\square:=\frac{1}{\sqrt{-g}}\partial_{\mu}\sqrt{-g}\partial^{\mu}$ and $g:=\text{det}(g_{\mu\nu})$.
To solve the Klein-Gordon equation \eqref{klein:gordon:equation} we first notice that every $1+1$ dimensional space-time is conformally flat \cite{Birrell:Davies,Wald:84}. This implies that there always exist coordinates $u=u(t,r),v=v(t,r)$ such that the Klein Gordon equation \eqref{klein:gordon:equation} takes the form
\begin{eqnarray}
\partial_u\partial_v\Phi(u,v)=0\label{eddington:finkelstein:klein:gordon:equation}.
\end{eqnarray}
In our case, we employ the Eddington-Finkelstein advanced and retarded coordinates $u,v$ defined by
\begin{eqnarray}
u&:=&ct-r_*\nonumber\\
v&:=&ct+r_*,\label{ef:coordinates}
\end{eqnarray}
where the tortoise coordinate $r_*$ is defined as $r_*:=r+r_S\ln|\frac{r}{r_S}-1|$ \cite{Misner:Thorne:73,Wald:84}.
Solutions to the Klein Gordon equation \eqref{eddington:finkelstein:klein:gordon:equation} can be expanded in terms of modes of the form
\begin{eqnarray}
\phi^{(u)}_{\omega}(u)&=&\frac{e^{i\omega u}}{2\sqrt{\pi\omega}}\nonumber\\
\phi^{(v)}_{\omega}(v)&=&\frac{e^{i\omega v}}{2\sqrt{\pi\omega}},\label{mode:solution}
\end{eqnarray}
which represent outgoing and ingoing waves that follow geodesics $u=$const and $v=$const, respectively. The frequency $\omega>0$ is the frequency as measured by an observer (infinitely) far from the planet with respect to his proper time $t$. Furthermore, the mode solutions \eqref{mode:solution} are eigenfunctions of the timelike Killing vector $i\partial_t$ \cite{Birrell:Davies} and therefore satisfy the eigenvalue equation
\begin{eqnarray}
i\partial_t\phi^{(u)}_{\omega}&=&\omega\phi^{(u)}_{\omega}\nonumber\\
i\partial_t\phi^{(v)}_{\omega}&=&\omega\phi^{(v)}_{\omega}.\label{eigenvalue:equation}
\end{eqnarray}
The mode solutions \eqref{mode:solution} are normalized through the standard conserved inner product $(\cdot,\cdot)$ (see \cite{Srednicki}) by
\begin{eqnarray}
(\phi_{\omega}^{(u)},\phi_{\omega'}^{(u)})&=&(\phi_{\omega}^{(v)},\phi_{\omega'}^{(v)})=\delta(\omega-\omega')
\end{eqnarray}
while mixed inner products vanish.
Finally, the quantum field $\Phi$ can be expanded as
\begin{eqnarray}
\Phi=\int_0^{+\infty}d\omega\left[\phi_{\omega}^{(u)}a_{\omega}+\phi_{\omega}^{(v)}b_{\omega}+\text{h.c.}\right],
\end{eqnarray}
where the bosonic annihilation operators $a_{\omega},b_{\omega}$ annihilate the vacuum state $|0\rangle$ through the standard relation $a_{\omega}|0\rangle=b_{\omega}|0\rangle=0$ and satisfy the canonical commutation relations
\begin{eqnarray}
\left[a_{\omega},a^{\dagger}_{\omega'}\right]=\left[b_{\omega},b^{\dagger}_{\omega'}\right]=\delta(\omega-\omega'),\label{sharp:canonical:commutation:relations}
\end{eqnarray}
where mixed commutators vanish.

\section{Preparation, propagation and detection of photons\label{photons}}
We assume ideal optical sources, and will study the propagation on a curved background and how they are affected by such propagation. In particular, we are interested in finding a transformation between the frequency distribution of an optical mode as measured locally before and after propagation.

\subsection{Preparation}
In general, a photon can be modeled by a wave packet with a distribution $F(\omega)\in\mathbb{C}$ peaked around a central frequency $\omega_0$ \cite{Downes:Ralph:13,Leonhardt}. The annihilation operator for the photon which, for an observer (infinitely) far from the planet, takes the form
\begin{eqnarray}
a_{\omega_0}(t)=\int_0^{+\infty}d\omega\, e^{-i\omega t} F_{\omega_0}(\omega)\,a_{\omega}\label{schwarzschild:wave:packet}.
\end{eqnarray}
The photon creation and annihilation operators $a^{\dagger}_{\omega_0},a_{\omega_0}$ satisfy the canonical equal time bosonic commutation relations
\begin{eqnarray}
\left[a_{\omega_0}(t),a^{\dagger}_{\omega_0}(t)\right]=1
\end{eqnarray}
if the frequency distribution $F(\omega)$ is normalized, i.e. $\int_{\omega>0} d\omega |F(\omega)|^2=1$. Such a distribution naturally arises if the optical field is described as a spatially and temporally localized propagating physical system \cite{Bruschi:Lee:13} i.e., a pulse.

Let Alice and Bob be two observers sitting at different constant distances from the surface of the planet. We can assume that Alice has her laboratory on the surface, $r_A=r_E$, while Bob has his lab on a satellite at constant distance $r_B$ from the surface, $r_B>r_A$.
Alice and Bob measure frequencies in their laboratories \textit{with respect to their clocks}, i.e., with respect to their proper times $\tau_A$ and $\tau_B$. By employing the definition of proper time \eqref{proper:time} and the eigenvalue equation \eqref{eigenvalue:equation}, it is simple to show that \eqref{eigenvalue:equation} is equivalent to
\begin{eqnarray}
i\partial_{\tau_K}\phi^{(u)}_{\Omega_K}&=&\Omega_K\phi^{(u)}_{\Omega_K},\label{proper:time:eigenvalue:equation}
\end{eqnarray}
where $K=A,B$ labels Alice or Bob and analogous formulas hold for $\phi^{(v)}$. In \eqref{proper:time:eigenvalue:equation}, we have introduced the physical frequency $\Omega_K$ as measured by the observer at radius $r_K$ as
\begin{eqnarray}
\Omega_K=\frac{\omega}{\sqrt{f(r_K)}}.
\end{eqnarray}
Since $\omega t$ is observer independent, if Alice prepares a sharp frequency mode $\Omega_A$, Bob will receive the frequency
\begin{eqnarray}
\Omega_B=\sqrt{\frac{f(r_A)}{f(r_B)}}\Omega_A,\label{physical:frequency:relation}
\end{eqnarray}
which is the well known formula for gravitational redshifts \cite{Misner:Thorne:73}. It is immediate to show that the relation between $\tau_B$ and $\tau_A$ is
\begin{eqnarray}
\tau_B=\sqrt{\frac{f(r_B)}{f(r_A)}}\tau_A.\label{proper:time:relation}
\end{eqnarray}
In real implementations, special relativistic effects might also affect our systems as satellites might not follow geo-stationary orbits. In this case, satellites will have a velocity component with respect to observers on the ground. It is well known from special relativity that frequencies emitted and received by two observers in relative (uniform) motion are doppler-shifted. 

%\hl{A more detailed analysis is needed to compute the exact impact of motion on the shifts in frequency. This would require solving the field equations in at least $2+1$ dimensions (since satellite orbits lie on a plane). As pointed out before, there are no analytical solutions to the field equations in more than $1+1$ dimensions and it is reasonable to assume that the angular component of the modes would not significantly contribute to the purely gravitational effects we are interested in. Nevertheless, we can exactly compute in $3+1$ dimensions what is the \textit{total} relativistic frequency shift between modes emitted by a source with four velocity $u_{A}^{\mu}$ and the one measured by a receiver with four velocity $v_{B}^{\mu}$.

A more detailed analysis is needed to compute the exact impact of motion on the shifts in frequency. This would require solving the field equations in at least $2+1$ dimensions (since satellite orbits lie on a plane). As pointed out before, there are no analytical solutions to the field equations in more than $1+1$ dimensions and it is reasonable to assume that the angular component of the modes would not significantly contribute to the purely gravitational effects we are interested in. The angular part contributes mainly to spreading the beam, an effect which is not of interest here. Nevertheless, we can exactly compute, in $3+1$ dimensions, the \textit{total} relativistic frequency shift between modes emitted by a source and the ones measured by a receiver.

%The source, Alice, with four-velocity $u_{A}^{\mu}$ sends an electromagnetic wave to an observer, Bob, whose four-velocity is $v_{B}^{\mu}$. We assume that Bob and Alice are both on a circular. Let $w^{\nu}$ be the tangent vector of the null geodesic that the light follows. Then the emission frequency $\Omega_A$ and the absorption frequency $\Omega_B$ are related by }
%\begin{equation*}
%\frac{\Omega_B}{\Omega_A}=\frac{\left.v^{\mu}w_{\mu}\right|_B}{\left.u^{\mu}w_{\mu}\right|_A}, 
%\end{equation*}
%which can can be readily computed using, for example, results in \cite{LEE THESIS}. One therefore finds 
%\begin{equation}
%\Omega_B=\sqrt{\frac{1 - \frac{2M}{r_A}}{1 - \frac{3M}{r_B}}}\Omega_A\label{total:physical:frequency:relation},
%\end{equation}
%\hl{which gives the total frequency shift for the $3+1$ dimensional case.}

Suppose, the source, Alice, with four-velocity $u_{A}^{\mu}$, in different dimensions $\mu$, sends an electromagnetic wave to an observer, Bob, whose four-velocity is $v_{B}^{\mu}$. We assume that Bob and Alice follow a (different) circular orbit. Let $w^{\nu}$ be the tangent vector of the null geodesic that the light follows. Then the emission frequency $\Omega_A$ and the absorption frequency $\Omega_B$ are related by 
\begin{equation*}
\frac{\Omega_B}{\Omega_A}=\frac{\left.v^{\mu}w_{\mu}\right|_B}{\left.u^{\mu}w_{\mu}\right|_A}, 
\end{equation*}
which can be further simplified using the results in \cite{Hodgkinson:13} to obtain 
\begin{equation}
\Omega_B=\sqrt{\frac{1 - \frac{2M}{r_A}}{1 - \frac{3M}{r_B}}}\Omega_A\label{total:physical:frequency:relation},
\end{equation}
which gives the total frequency shift in the $3+1$ dimensional case.

\subsection{Propagation}
In order to understand how the propagation of light is affected by the background space-time, we start by noting that Alice or Bob will describe the optical mode \eqref{schwarzschild:wave:packet} through the operator
\begin{eqnarray}
a_{\Omega_{K,0}}(\tau_K)=\int_0^{+\infty}d\Omega_K\, e^{-i\Omega_K \tau_K} F^{(K)}_{\Omega_{K,0}}(\Omega_K)\,a_{\Omega_K},\label{wave:packet}
\end{eqnarray}
where $K=A,B$ labels either Alice or Bob, $\Omega_K$ are the physical frequencies as measured in their labs using the proper times $\tau_K$ and $\Omega_{K,0}$ are the peak frequencies of the frequency distributions $F^{(K)}_{\Omega_{K,0}}$. The operators $a_{\Omega_K}$ must satisfy the canonical commutation relations
\begin{eqnarray}
[a_{\Omega_K},a^{\dagger}_{\Omega_K'}]=\delta(\Omega_K-\Omega^{'}_K).\label{sharp2:canonical:commutation:relations}
\end{eqnarray}
Our aim in this section is to find the relation between the shape of the wave packet $F^{(A)}_{\Omega_{A,0}}$ prepared by Alice at some time $\tau_A$ and the shape of the wave packet $F^{(B)}_{\Omega_{B,0}}$ received by Bob at some time $\tau_B>\sqrt{f(r_B)/f(r_A)}\tau_A$ after propagation through space-time.
It is important to notice that Alice's and Bob's operators \eqref{wave:packet} can be used to describe the \textit{same} optical mode in two \textit{different} frames before and after propagation.

We start by noting that outgoing photons follow geodesics of the form $u=$const. If a mode is (sharply) localized around $r_0$ at $t_0=0$, by using \eqref{ef:coordinates} one can show that at time $t>t_0$ it will be (sharply) localized around $r$ given implicitly by
\begin{eqnarray}
r_*(r)=r_*(r_0)+t.\label{propagation}
\end{eqnarray}
Equation \eqref{propagation} informs us about at which time Bob will detect photons. On the other hand, equations \eqref{physical:frequency:relation} and \eqref{proper:time:relation} inform us of the relation between the frequencies and proper times measured by Alice and those measured by Bob.
It is now necessary to find the relation between the operators $a_{\Omega_A}$ and $a_{\Omega_B}$. To do this we employ \eqref{sharp:canonical:commutation:relations}, \eqref{physical:frequency:relation}, \eqref{sharp2:canonical:commutation:relations} and the identity $\delta(f(x))=\sum_i\frac{\delta(x-x_{i})}{\left.|\frac{\partial f}{\partial x}|\right|_{x_{i}}}$, where $f(x_{i})=0\,\forall i$, to write
\begin{eqnarray}
\left[a_{\omega},a^{\dagger}_{\omega'}\right]=\frac{1}{\sqrt{f(r)}}\left[a_{\Omega},a^{\dagger}_{\Omega'}\right].
\end{eqnarray}
This implies that $a_{\Omega}=\sqrt[4]{f(r)}a_{\omega}$. Employing all of these relations, it is now easy to see that the operator \eqref{wave:packet} can be written before and after propagation as a function of measurable quantities in Alice or Bob's labs (i.e., peak frequency, bandwidth).
%as follows
%\begin{widetext}
%\begin{eqnarray}
%a_{\Omega_{A,0}}(\tau_A)&=&\int_0^{+\infty}d\Omega_A\, e^{-i\Omega_A \tau_A} F^{(A)}_{\Omega_{A,0}}(\Omega_A)\,a_{\Omega_A}\nonumber\\
%%=\sqrt[4]{\frac{f(r_B)}{f(r_A)}}\int_0^{+\infty}d\Omega_B\, e^{-i\Omega_B \tau_B} F^{(A)}_{\Omega_{A,0}}(\sqrt{\frac{f(r_B)}{f(r_A)}}\Omega_B)\,a_{\Omega_B}\nonumber\\
%a_{\Omega_{B,0}}(\tau_B)&=&\int_0^{+\infty}d\Omega_B\, e^{-i\Omega_B \tau_B} F^{(B)}_{\Omega_{B,0}}(\Omega_B)\,a_{\Omega_B},\label{packet}
%\end{eqnarray}
%\end{widetext}
%where $\tau_B>\sqrt{f(r_B)/f(r_A)}\tau_A$ and we
Finally, using \eqref{schwarzschild:wave:packet} as an intermediate step, we find that the frequency distributions $F^{(K)}_{\Omega_{K,0}}$ as measured in different reference frames satisfy the relation
\begin{eqnarray}
F^{(B)}_{\Omega_{B,0}}(\Omega_B)=\sqrt[4]{\frac{f(r_B)}{f(r_A)}}F^{(A)}_{\Omega_{A,0}}(\sqrt{\frac{f(r_B)}{f(r_A)}}\Omega_B).\label{wave:packet:relation}
\end{eqnarray}
Alice can prepare the wave packet $a_{\Omega_{A,0}}(\tau_A)$ in \eqref{wave:packet} at time $\tau_A$ and send it to Bob who receives it as $a_{\Omega_{B,0}}(\tau_B)$ at time $\tau_B>\sqrt{f(r_B)/f(r_A)}\tau_A$ which depends on the user's relative distance (i.e., equation \eqref{propagation}). Equation \eqref{wave:packet:relation} informs us that, in general, the wave packet received by Bob will have a different peak frequency and a different shape than those prepared by Alice. In particular, for the scenario of interest where Bob finds himself at higher altitudes than Alice ($r_B>r_A$), the wave packet frequencies $\Omega_{B}$ as measured by Bob will all be redshifted with respect to those as created by Alice (see \eqref{physical:frequency:relation}).

Notice that if we wish to take into account the full effects of gravity and motion of the satellite, we find the updated version of \eqref{wave:packet:relation} as 
\begin{eqnarray}
F^{(B)}_{\Omega_{B,0}}(\Omega_B)=\sqrt[4]{\frac{1 - \frac{2M}{r_A}}{1 - \frac{3M}{r_B}}}F^{(A)}_{\Omega_{A,0}}(\sqrt{\frac{1 - \frac{2M}{r_A}}{1 - \frac{3M}{r_B}}}\Omega_B).\label{total:wave:packet:relation}
\end{eqnarray}

We emphasize that the effect described by \eqref{wave:packet:relation} cannot be simply corrected by a linear shift of frequencies. Therefore, it may be challenging to compensate the transformation induced by the curvature in realistic implementations.

\subsection{Detection}

Before leaving Earth to reach his station at height $r_B$, Bob agreed with Alice to communicate using light described by the wave packet $F^{(A)}_{\Omega_{A,0}}$. Let Alice prepare a pulse described by the mode operator $a_{\Omega_{A,0}}(\tau_A)$ which she then sends to Bob who receives it as $a_{\Omega_{B,0}}(\tau_B)$. We have shown that, in general, the wave packet $F^{(B)}_{\Omega_{B,0}}$ will be \textit{different} compared to the one Bob was expecting. The difference can be observed in equation \eqref{wave:packet:relation}. Regardless of the specific model of the detector, if the measuring device is tuned to click when a photon in the wave packet $F^{(A)}_{\Omega_{A,0}}$ is received, the probability of the detector to click when $F^{(B)}_{\Omega_{B,0}}$ is received will be affected.
Bob therefore will believe that the channel between him and Alice (i.e., the space-time) is noisy. He can quantify the ``goodness'' of the channel by employing the fidelity $\mathcal{F}$ defined as
\begin{eqnarray}
\mathcal{F}:=\text{Tr}^2\left[\sqrt{\sqrt{\rho}\rho^{\prime}\sqrt{\rho}}\right]\label{mixed:state:fidelity},
\end{eqnarray}
for arbitrary input states $\rho,\rho^{\prime}$. In case the input states are pure, for example  $\rho=|\psi\rangle\langle\psi|$ and $\rho^{\prime}=|\psi^{\prime}\rangle\langle\psi^{\prime}|$, the fidelity \eqref{mixed:state:fidelity} reduces to $\mathcal{F}=|\langle\psi|\psi^{\prime}\rangle|^2$ and the intensity fidelity gives the probability that the state was $\rho$ given that $\rho^{\prime}$ is obtained in a measurement.

\section{Transmission and reception of a single mode\label{photons:traveling}}
\begin{figure}
\includegraphics[width=\linewidth]{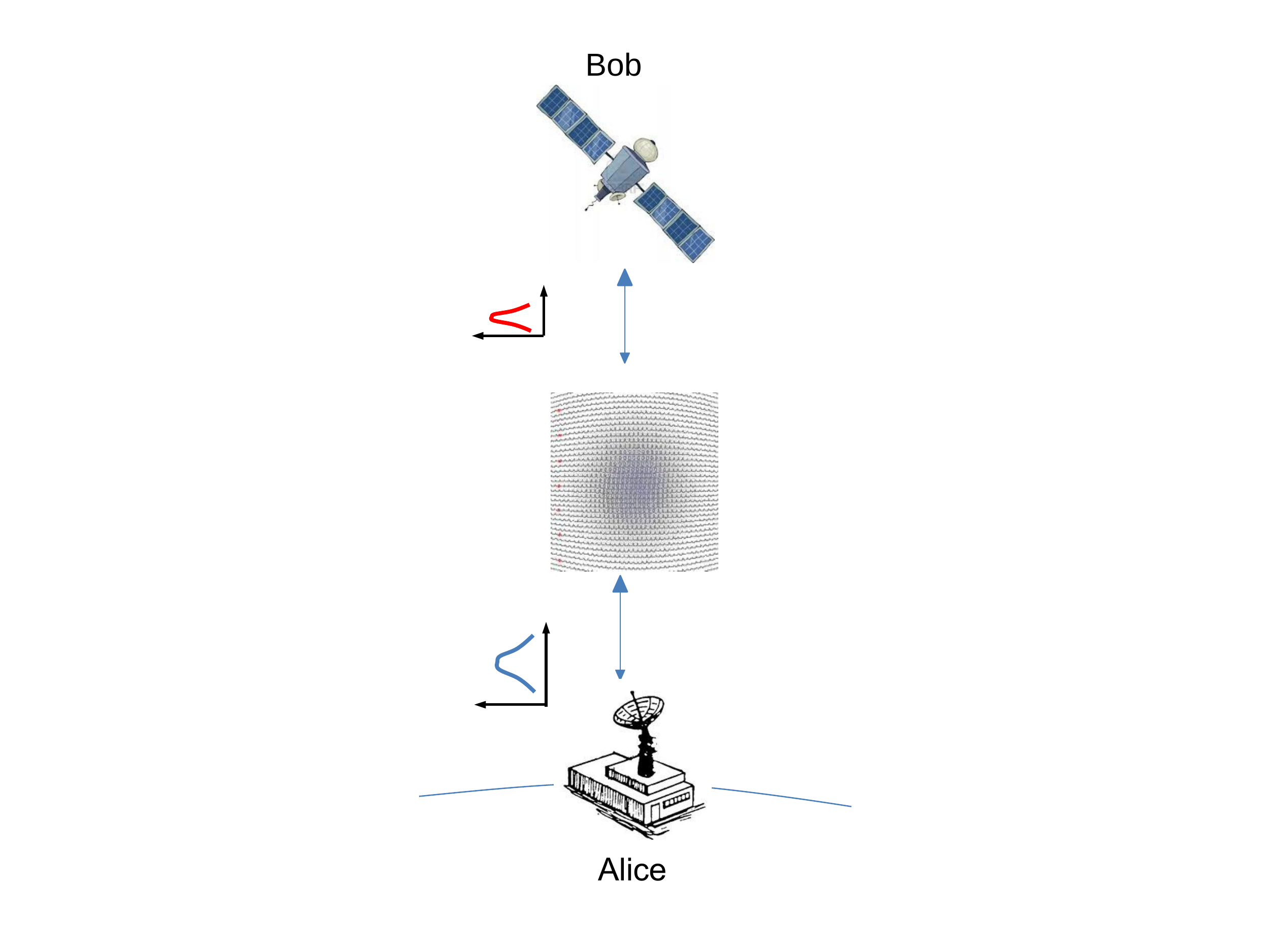}
\caption{\label{fig:setup} Illustration of the setup considered here. Alice and Bob are located at different heights in a gravitational potential. Alice prepares and sends photons (in this work  a single photon or a laser beam) to Bob, who uses them to complete a communication protocol. The photon is created by Alice with certain characteristics (i.e., peak frequency and bandwidth), which change once it is received by Bob.}
\end{figure}

Alice and Bob can communicate using a wealth of protocols \cite{Scarani:BechmannPasquinucci:09}. In order to illustrate the techniques developed in the previous section we start with a few simple examples before moving on to more realistic communication schemes. Here we analyze the transmission of a single photon, then the transmission of a coherent state and finally the transmission of a mode which is part of a two mode squeezed state (the scheme is illustrated in figure \ref{fig:setup}).

\subsection{Single photon}
Conceptually, the simplest protocol is when Alice prepares a single photon in the mode $F^{(A)}_{\Omega_{A,0}}$ and sends it to Bob. The state $|\psi_{\text{s.p.}}\rangle$ of the system at time $\tau_A=0$ is
\begin{eqnarray}
|\psi_{\text{s.p.}}\rangle=a^{\dagger}_{\Omega_{A,0}}(0)|0\rangle.
\end{eqnarray}
Bob now receives the photon
\begin{eqnarray}
|\psi_{\text{s.p.}}\rangle=a^{\dagger}_{\Omega_{B,0}}(0)|0\rangle,
\end{eqnarray}
which is characterized by a distribution $F^{(B)}_{\Omega_{B,0}}(\Omega_B)$ different from the distribution $F^{(A)}_{\Omega_{A,0}}(\Omega_B)$ that Alice promised to send him. The intensity fidelity $\mathcal{F}^2$ depends on the fidelity $\mathcal{F}_{\text{s.p.}}$ of the channel, which in this case is simply the overlap of the two distributions
\begin{eqnarray}
\mathcal{F}_{\text{s.p.}}=\left|\Delta\right|^2\label{single:photon:fidelity},
\end{eqnarray}
where
\begin{eqnarray}
\Delta:=\int_0^{+\infty}d\Omega_B\,F^{(B)\star}_{\Omega_{B,0}}(\Omega_B)F^{(A)}_{\Omega_{A,0}}(\Omega_B).\label{single:photon:fidelity}
\end{eqnarray}
Clearly $\Delta=1$ for a perfect channel. If the curvature is strong enough, the distributions in \eqref{single:photon:fidelity} might have negligible overlap and the fidelity would be low. In the case of Earth-to-LEO communication, we will show that the fidelity \eqref{single:photon:fidelity} is $|\Delta|^2\sim1-2\times10^{-11}$.

\subsection{Coherent state}
Alice now decides to send Bob a laser pulse instead of a single photon. An initial coherent state $|\alpha\rangle$ prepared by Alice with displacement $\alpha$ takes the form
\begin{eqnarray}
\bigl|\alpha\rangle=\left.\hat{D}_{A}(\alpha)\bigl|0\rangle\right|_{\tau_{0,A}=0},
\end{eqnarray}
where the \textit{displacement operator} is defined as $\hat{D}(\alpha)(\tau_A):=\exp(\alpha a^{\dagger}_{\Omega_{A,0}}(\tau_A)-\alpha^*a_{\Omega_{A,0}}(\tau_A))$. Bob will receive a coherent state with the same displacement parameter $\alpha$ but defined for different modes $F^{(B)}_{\Omega_{B,0}}$. The fidelity $\mathcal{F}_{\text{c.s.}}$ in this case will read
\begin{eqnarray}
\mathcal{F}_{\text{c.s.}}=e^{-2|\alpha|^2(1-\Re(\Delta))},
\end{eqnarray}
where $\Re(\Delta)$ denotes the real part of $\Delta$.

\subsection{Two mode squeezed state}
As a last case Alice will send Bob one mode of two mode squeezed state $|s\rangle$. The state $|s\rangle$ takes the form
\begin{eqnarray}
\bigl|s\rangle=\left.\hat{S}_{A}(\alpha)\bigl|0\rangle\right|_{\tau_{0,A}=0},\label{two:mode:squeezed:state}
\end{eqnarray}
where the \textit{squeezing operator} is defined as
\begin{eqnarray*}
\hat{S}(s)(\tau_A):=e^{ s\left[ a^{\dagger}_{\Omega_{A,0}}(\tau_A)b^{\dagger}_{\Omega^{\prime}_{A,0}}(\tau_A)-a_{\Omega_{A,0}}(\tau_A)b_{\Omega^{\prime}_{A,0}}(\tau_A)\right]}
\end{eqnarray*}
and $s$ is known as the squeezing parameter \cite{Leonhardt,Scully:Zubairy:02}. We assume that Alice has prepared the mode $b_{\Omega^{\prime}_{A,0}}(\tau_A)$ which is received as mode $b^{\prime}_{\Omega^{\prime}_{B,0}}(\tau_B)$ by Bob. In this scenario, an operational definition of fidelity involves comparing the state $|s\rangle$ that Bob and Alice expect to share with the state $|s\rangle^{\prime}$ they actually share after the propagation of mode $b_{\Omega^{\prime}_{A,0}}$. The fidelity computed this way sets a lower bound to the average fidelity of communication between Alice and Bob. The state $|s\rangle^{\prime}$ takes the form \eqref{two:mode:squeezed:state} with $b^{\prime}_{\Omega^{\prime}_{A,0}}$ in place of $b_{\Omega^{\prime}_{A,0}}$.
We can compute the fidelity $\mathcal{F}_{\text{t.s.}}$ for this scenario as $\mathcal{F}_{\text{t.s.}}=|\langle s|s\rangle^{\prime}|$ and we obtain
\begin{eqnarray}
\mathcal{F}_{\text{t.s.}}=\left|\frac{1}{\cosh^2 s}\frac{1}{1-\Delta\tanh^2 s}\right|^2.\label{tmss:fidelity}
\end{eqnarray}
Note that, no matter how well the modes overlap (i.e., how small is $1-|\Delta|$) as long as the overlap is not perfect the fidelity \eqref{tmss:fidelity} vanishes for infinite squeezing.

\section{Communication between different altitudes (or communication in a gravitational potential)\label{protocol}}

\subsection{Establishing entangled links for communication protocols}
Communication protocols based on discrete variables often require two users to employ and exchange qubits and share a maximally entangled state \cite{Ekert:91,Bennett:Brassard:92,Bennett:Brassard:93}. A qubit can, for example, be  physically implemented by the two polarization states of a photon. There are many protocols that enable two distant users to share entangled states \cite{Duan:Lukin:01,Razavi:Shapiro:06,Sangouard:Simon:07,Sangouard:Simon:11}, but Alice and Bob may wish to employ a protocol that requires only local operations. Alice and Bob can obtain an entangled state between two memories by entangling them locally with two photons and then performing a Bell state measurement on the state of the two photons. This is the same idea as used in measurement-device-independent QKD \cite{Sangouard:Simon:07,HoiKwong:Curty:12,Braunstein:Pirandola:12}, where Alice and Bob use entanglement swapping to share a key. In either case, this requires at least one of the two photons to travel through space-time and we have previously demonstrated that photons will in general be affected.

\subsubsection{Flat space-time entangling protocol\label{protocol:section}}
\begin{figure}
\includegraphics[width=\linewidth]{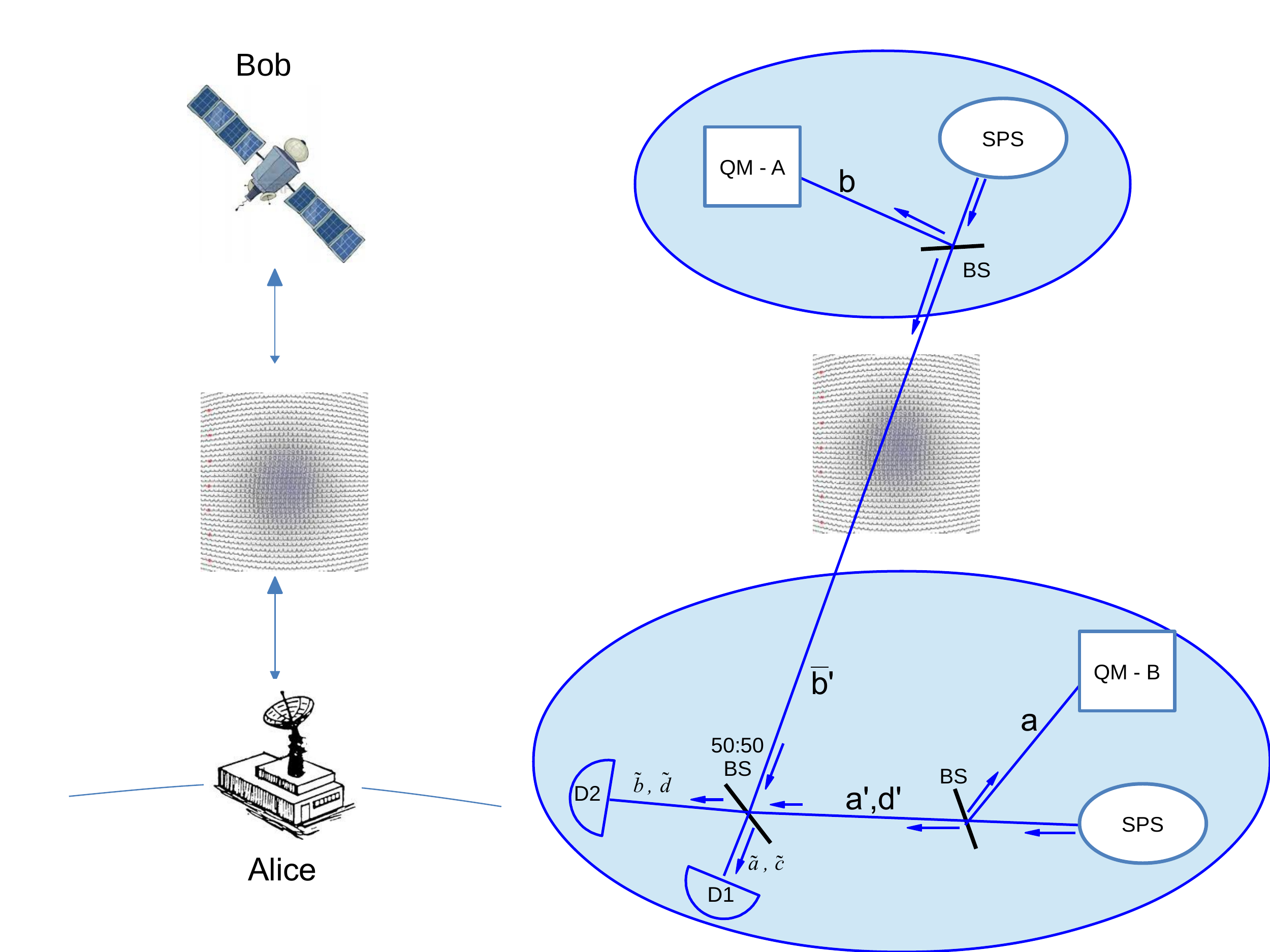}
\caption{\label{fig:SPS}(Color online) Schematic diagram for entanglement
distribution between two quantum memories (QMs) located at Alice's and Bob's locations. Single photon sources (SPSs), memories and detectors are represented by circles, squares and half-circles, respectively. Vertical bars denote beam splitters. In this protocol, the detection of a single
photon after a beam splitter ideally projects the two memories onto an entangled state.}
\end{figure}

We briefly describe the ideal flat space-time setup employed by our users.
The scheme that Alice and Bob will use is depicted in Fig.~\ref{fig:SPS}. Alice and Bob have a quantum memory and a single photon source each. The single photon sources produce one photon each, which travel through two balanced (for simplicity) beam splitters located at the respective stations, as shown in Fig.~\ref{fig:SPS}. The photons are either transmitted (modes $a^{\prime}b^{\prime}$) or are reflected and are then stored in the memory (modes $a,b$). The state $\bigl|\Psi_{\text{in}}\rangle$ of the system at this point of the scheme is
\begin{eqnarray}
\bigl|\Psi_{\text{in}}\rangle=\frac{1}{2}\left[\bigl|\underset{aba^{\prime}b^{\prime}}{\underbrace{1100}}\rangle+\bigl|0110\rangle+\bigl|1001\rangle+\bigl|0011\rangle\right].
\end{eqnarray}
The modes $a^{\prime},b^{\prime}$ are then recombined at a second balanced beam splitter located at the Alice's station and the output modes $\tilde{a},\tilde{b}$ are measured. Here, we assumed the phase difference between the two paths is fixed and compensated. Time synchronization is also in place to ensure that the photons arrive at the same time at the beam splitter of the measurement module. Note that the overall phase of the initial single photons does not affect the final state obtained and no coordination is needed to drive both single-photon sources coherently.

To understand what is the action of the detection process we start by writing the transformation between modes $a^{\prime},b^{\prime}$ and modes $\tilde{a},\tilde{b}$ as
\begin{align}
\begin{pmatrix}
\tilde{a}\\
\tilde{b}
\end{pmatrix}
=
\frac{1}{\sqrt{2}}
\begin{pmatrix}
1 & 1\\
1 & -1
\end{pmatrix}
\begin{pmatrix}
a^{\prime}\\
b^{\prime}
\end{pmatrix}.\label{beam:splitter}
\end{align}
If modes $\tilde{a}$ or $\tilde{b}$ are detected, i.e. detectors D1 or D2 click, the state of the memories is projected respectively into
\begin{eqnarray}
\rho_{\pm}=\frac{1}{2}\mathbf{P}_{\pm}+\frac{1}{2}\mathbf{P}_{\text{vac.}}
\end{eqnarray}
where $\mathbf{P}_{\pm}$ denote projectors on the maximally entangled states
\begin{eqnarray}
\bigl|\Psi_{\pm}\rangle=\frac{1}{\sqrt{2}}\bigl[\bigl|\underset{ab}{\underbrace{10}}\rangle\pm\bigl|01\rangle\bigr],
\end{eqnarray}
while $\mathbf{P}_{\text{vac.}}$ denotes the projection on the vacuum state. In case number resolving photodetectors are employed, we can exclude the cases when two photons arrive at the same detector and therefore ideally $\bigl|\Psi_{\text{out}}\rangle=\bigl|\Psi_{\pm}\rangle$, with $|\phi \rangle^+ (|\phi \rangle^-)$ heralded by a click at D1 (D2). Note that there is still a chance that one of the two photons is lost along the way, therefore degrading the state. For the sake of our argument, in this work we are not interested in any source of loss or imperfection and from now on we will assume that detectors are ideal and photons always reach their destination.

\subsubsection{Curved spacetime entangling protocol\label{curved:spacetime:protocol:section}}
The scheme described above works when Alice and Bob are in flat space-time. Let Alice and Bob now be in curved Schwarzschild space-time in the same setup as described in the previous section and depicted in figure \ref{fig:SPS}. We assume for simplicity that Alice's modes act as reference modes, since it is much more feasible that all operations of interest occur at her station. In case the operations were performed between Alice and Bob or at Bob's station, the results would be qualitatively the same. Bob will generate a photon in mode $b^{\prime}$ which  will enter the beam splitter on Earth as a different mode $\bar{b}^{\prime}$ than the expected one. Therefore, we can decompose the mode $\bar{b}^{\prime}$ that reaches Earth as
\begin{eqnarray}
\bar{b}^{\prime}=\sqrt{1-q}\,b^{\prime}+\sqrt{q}\,c^{\prime},\label{mode:decomposition}
\end{eqnarray}
where $q\leq1$. Equation \eqref{mode:decomposition} states that when Bob's mode reaches the Earth, it will have a contribution from mode $b^{\prime}$ (which matches Alice's mode $a^{\prime}$) and a contribution from the orthogonal mode $c^{\prime}$, i.e. $[a^{\prime},c^{\prime\dagger}]=0$. Such a decomposition is always possible \cite{Rohde:Mauerer:07}.

The parameter $q$ is directly related to the fidelity of single photon transmission as defined in \eqref{single:photon:fidelity}. In fact,
\begin{eqnarray}
\Delta=\langle0\bigr|\bar{b}^{\prime}b^{\prime\dagger}\bigl|0\rangle=\sqrt{1-q},
\end{eqnarray}
which implies
\begin{eqnarray}
q=1-\Delta^2.\label{overlap:fidelity:relation}
\end{eqnarray}
We can assume without loss of generality that $\Delta\in\mathbb{R}$. It is straightforward to generalize all of the following results to the case of complex $\Delta$.

The balanced beam splitter will mix mode $a^{\prime}$ with $b^{\prime}$ and mode $c^{\prime}$ with a corresponding mode $d^{\prime}$. Both couples are transformed to new outputs $\tilde{a},\tilde{b}$ and $\tilde{c},\tilde{d}$ respectively by a transformation of the form \eqref{beam:splitter}. Notice that since all operations occur at Alice's station, the mode $d^{\prime}$ is always in its ground state. The initial state $\bigl|\Psi_{\text{in}}\rangle$ of the whole system  before modes $a^{\prime}b^{\prime}c^{\prime}d^{\prime}$ enter the beam splitter on Earth is
\begin{widetext}
\begin{eqnarray}
\bigl|\Psi_{\text{in}}\rangle=\frac{1}{2}\left[\bigl|\underset{aba^{\prime}b^{\prime}c^{\prime}d^{\prime}}{\underbrace{110000}}\rangle+\bigl|011000\rangle+\sqrt{1-q}\bigl|100100\rangle+\sqrt{q}\bigl|100010\rangle+\sqrt{q}\bigl|001100\rangle+\sqrt{1-q}\bigl|001010\rangle\right].\label{new:initial:state}
\end{eqnarray}
\end{widetext}
 We can now invert the relation between $a^{\prime},b^{\prime}$ and $\tilde{a},\tilde{b}$ (analogously for the second couple of modes) and express the state \eqref{new:initial:state} in terms of memory modes $a,b$ and the beam splitter output modes $\tilde{a},\tilde{b},\tilde{c},\tilde{d}$. The expression involves fifteen terms and is not illuminating.
The Bell state measurement on Earth is completed when the photodetectors absorb the photons.
If detector D1 clicks, modes $\tilde{a}$ and $\tilde{c}$ have been detected. The projection operator that implements this detection is
\begin{eqnarray}
\mathbf{D}_+=\mathds{1}_{a,b}\otimes\hat{N}\otimes|0\rangle\langle0|\otimes \hat{N}\otimes|0\rangle\langle0|,
\end{eqnarray}
where $\hat{N}:=\mathds{1}-|0\rangle\langle0|$ and the order of the modes is $a,b,\tilde{a},\tilde{b},\tilde{c},\tilde{d}$. If detector D2 clicks, this corresponds to projecting the initial state $\bigl|\Psi_{\text{in}}\rangle$ with the operator $\mathbf{D}_-$ of the form
\begin{eqnarray}
\mathbf{D}_-=\mathds{1}_{a,b}\otimes|0\rangle\langle0|\otimes\hat{N}\otimes|0\rangle\langle0|\otimes\hat{N}.
\end{eqnarray}
The final states
\begin{eqnarray}
\rho_{\pm}(q):=\frac{\text{Tr}_{\text{phot.}}[\mathbf{D}_{\pm}|\Psi_{\text{in}}\rangle \langle\Psi_{\text{in}}|]}{\text{Tr}[\mathbf{D}_{\pm}|\Psi_{\text{in}}\rangle \langle\Psi_{\text{in}}|]}
\end{eqnarray}
of the memories, after detection and absorption of single photons by the resolving photodetectors, are respectively
\begin{eqnarray}
\rho_{\pm}=\frac{1}{2}\left[(1\pm\sqrt{1-q})\mathbf{P}_{+}+(1\mp\sqrt{1-q})\mathbf{P}_{-}\right].\label{final:state:memories}
\end{eqnarray}
In equation \eqref{final:state:memories}, $\mathbf{P}_{\pm}$ represents the projector on the state $|\Psi_{\pm}\rangle$.
Since $\rho_{\pm}(q)$ is a mixed state, we can compute the Negativity $\mathcal{N}$ which is a measure of entanglement based on the positivity of the partial transpose (PPT criterion) \cite{Peres:96,Vidal:Werner:2002}. In order to compute the negativity we first need to find the partial transpose $\rho^{\text{PT}}_{\pm}$ of the state $\rho_{\pm}$. This can be obtained by transposing the subspace of one of the two modes. Then, we define
\begin{eqnarray}
\mathcal{N}[\rho^{\text{PT}}_{\pm}]:=\text{max}\left\{0,\sum_i \frac{|\lambda_i|-\lambda_i}{2}\right\},\label{negativity}
\end{eqnarray}
where $\lambda_i$ are the eigenvalues of the partial transposed state $\rho^{\text{PT}}_{\pm}$.
In our scenario it is easy to show that
\begin{eqnarray}
\mathcal{N}=\frac{\sqrt{1-q}}{2}.\label{scenario:negativity}
\end{eqnarray}
The Negativity reaches the value $\mathcal{N}=1/2$ for maximally entangled states. It is easy to see from \eqref{mode:decomposition} that when $q=0$ the mode sent by Bob reaches Alice intact (i.e., center and shape of the frequency distribution is unchanged) and therefore the memories are projected into a maximally entangled state as expected (see \eqref{final:state:memories}).
%We anticipate that our results in section \ref{results} will show that the negativity can be affected up to $\sim1\%$.

\subsection{A simple continuous-variable QKD protocol}
As an instructive counter example we now consider a type of quantum communication protocol whose performance is not affected by curvature. While some protocols require the users to perform only local operations and exchange quantum systems, other protocols require the users to exchange additional systems, such as local oscillators. The latter systems might naturally incorporate the means for the compensation of the effects due to space-time curvature. Here,
we will use the techniques developed above to analyze one such example of a continuous variable QKD protocol similar to that investigated in \cite{Downes:Ralph:13}. Alice employs two coherent states originating from the same source (i.e. a laser) with strong power. The source is then split up into one strong beam which is used as a reference (the Local Oscillator) and one weak beam used as a signal. Bob collects the two beams, mixes them at a balanced beam splitter and measures the photocurrents of the two output modes. In this section we will show that by using extra resources (i.e., the Local Oscillator) Bob will be able to compensate for the effects of the curvature of space-time.

Let Alice prepare the signal and the LO initially in two different coherent states $|\alpha\rangle$ and $|\beta\rangle$ of modes $a_{\text{S,A}}$,$a_{\text{L,A}}$ with displacement parameters that satisfy $\beta\gg|\alpha|$. Since the  modes come from the same source, they have the same frequency distribution ($\mathcal{F}_{\Omega_{L,A}}=\mathcal{F}_{\Omega_{S,A}}=\mathcal{F}_{\Omega_{0,A}}$). A coherent state is obtained by acting with the displacement operator $\hat{D}(\gamma)=\exp({\gamma\hat{a}^{\dag}-\gamma^*\hat{a}})$ on the vacuum state $|0\rangle$.
The initial state for this kind of protocol is therefore
\begin{eqnarray}
\bigl|\psi_i\rangle=\left.\hat{D}_{S,A}(\alpha)\hat{D}_{L,A}(\beta)\bigl|0\rangle\right|_{\tau_{0,A}=0}\label{initial:state:qkd:protocol},
\end{eqnarray}
where the subscripts $S,L$ denote the signal or the local oscillator and subscripts $A,B$ the user that prepares and/or receives it.
The modes $a_{\text{S,A}}$ and $a_{\text{L,A}}$ propagate and reach Bob, who mixes them at a balanced beam splitter described by the transformation  \eqref{beam:splitter}. We know that the modes received by Bob are \textit{different} from the modes sent by Alice, i.e., the peak frequency and the width of the distributions are different as measured by Alice and Bob. Bob will perform a measurement, i.e., count incident photons, for some time much longer than the bandwidths considered in this problem (see \cite{Downes:Ralph:13}). Therefore, he will integrate the input signals of his detectors over an infinite (proper) time.
The operator that describes the outcomes of balanced homodyne detection  at Bob's satellite (and with respect to \textit{his reference frame}) is \cite{Downes:Ralph:13,Scully:Zubairy:02,Leonhardt}
\begin{eqnarray}
\hat{O}:=\int_{-\infty}^{+\infty}d\tau_B\,\left[\hat{a}_{S,B}^{\dag}(\tau_B)\hat{a}_{L,B}(\tau_B)+\text{h.c.}\right]\label{homodyne:operator}.
\end{eqnarray}
Bob will compute the expectation value of such operator using the state $\bigl|\psi_i\rangle$ he receives. We are interested in the final expectation value $X:=\langle\psi_i\bigr|\hat{O}\bigl|\psi_i\rangle$.
%\begin{eqnarray}
%X&:=&\langle\psi_i\bigr|\hat{O}\bigl|\psi_i\rangle\label{observable}
%%&=&\int_{-\infty}^{+\infty}d\tau_B\,\langle\psi_i\bigr|\left[\hat{a}_{S,B}^{\dag}(\tau_B)\hat{a}_{L,B}(\tau_B)+\hat{a}_{S,B}(\tau_B)\hat{a}_{L,B}^{\dag}(\tau_B)\right]\bigl|\psi_i\rangle.\label{observable}
%\end{eqnarray}
In order to compute the observable X we follow \cite{Downes:Ralph:13} and assume that the detector is well localized in space and time. This implies that it responds with the same strength to a very broad range of frequencies, therefore $\hat{a}_{L,B}(\tau_B)$ and $\hat{a}_{S,B}(\tau_B)$ are broadband. We can commute the displacement operators through the mode operators as expressed in Bob's coordinates to give the following relations at Bob's site \cite{Downes:Ralph:13}
\begin{widetext}
\begin{eqnarray}
\hat{D}^{\dag}_{L,B}(\beta)\hat{a}_{L,B}(\tau_B)\hat{D}_{L,B}(\beta)&=&\left[\hat{a}_{L,B}(\tau_B)+\frac{\beta}{\sqrt{2\pi \Omega_0}}\int_{0}^{+\infty}d\Omega_B\, e^{-i\Omega_B \tau_B}\mathcal{F}_{\Omega_{0,B}}(\Omega_B)\right]\nonumber\\
\hat{D}^{\dag}_{S,B}(\alpha)\hat{a}_{S,B}(\tau_B)\hat{D}_{S,B}(\alpha)&=&\left[\hat{a}_{S,B}(\tau_B)+\frac{\alpha}{\sqrt{2\pi \Omega_0}}\int_{0}^{+\infty}d\Omega_B\, e^{-i\Omega_B \tau_B}\mathcal{F}_{\Omega_{0,B}}(\Omega_B)\right],\label{relations:bob}
\end{eqnarray}
\end{widetext}
where the signal/local oscillator modes he receives have the same frequency distribution $\mathcal{F}_{\Omega_{0,B}}$ as previously discussed.
Using the relations \eqref{relations:bob} we obtain $X$
\begin{eqnarray}
X&=&\beta\left[\alpha^*+\alpha\right].\label{ideal:result}
\end{eqnarray}
Another quantity of interest is the variance $V$ of this expectation value defined as \cite{Downes:Ralph:13,Scully:Zubairy:02,Leonhardt} $
V:=\langle\psi_i\bigr|\hat{O}^2\bigl|\psi_i\rangle-(\langle\psi_i\bigr|\hat{O}\bigl|\psi_i\rangle)^2$.
%\begin{eqnarray}
%V:=\langle\psi_i\bigr|\hat{O}^2\bigl|\psi_i\rangle-(\langle\psi_i\bigr|\hat{O}\bigl|\psi_i\rangle)^2\label{variance}.
%\end{eqnarray}
We can compute the variance $V$ and find
\begin{eqnarray}
V&=&2\left(|\beta|^2+|\alpha|^2\right)\sim 2|\beta|^2\label{variance:ideal:result}
\end{eqnarray}
since $\beta\gg|\alpha|$.

The result of the homodyne detection \eqref{ideal:result} and its variance \eqref{variance:ideal:result} \textit{are not} affected by the space-time curvature. One way to understand this conclusion is that Alice sends Bob a signal, which will change its frequency distribution profile, and a LO, which will be affected in the same way. Bob will use the LO as a reference beam for matching and detection of the input signal. Therefore, the effects of the change in frequency profile are compensated. We conclude that the key rate of any protocol using such quantum communication scheme will not be affected.

\section{Estimation of effects of space-time curvature on Earth-to-LEO quantum communication implementations}\label{results}
There has been extensive research on expanding the distances of quantum communications and QKD. For ground based systems, the hard limit is optical losses in fibers and free-space, which scales exponentially with distance \cite{Scarani:BechmannPasquinucci:09}. Quantum Repeaters are one way to extend distances on the ground, however there are still many fundamental challenges to be researched before they can be practical \cite{Sangouard:Simon:11}. Satellite transmissions solve this problem, because the transmission losses in empty space scale only quadratically with the distance. ÊWith todays technologies, distances of up to 100,000 km are feasible in empty space.

There are several international developments for satellite systems, all based on single photon systems using discrete variables: the Canadian researchers lead by Thomas Jennewein embarked in the mission QEYSSat (Quantum Encryption and Science Satellite) \cite{Rideout:Jennewein:12}, researchers in the USA within the research group of Richard Hughes and Jane Nordholt \cite{Buttler:Hughes:00}, the European groups headed by Anton Zeilinger in Vienna ( Space-QUEST project) \cite{Scheidel:Wille:13,Ma:Herbst:12,Scheidel:Wille:13},  Japanese researchers within the Japanese Space Agency as well as the National Institution of Information and Communication Technology (NICT) and the Chinese Academy of Science, which has announced openly that it is investigating the possibility of performing quantum communications in space (launch date of 2016) \cite{Hughes:Nordholt:11}.

The growing interest in developing and implementing efficient quantum networks in space motivates the estimation of all possible effects that can influence the reliability of the networks and jeopardize the missions. We have shown that entanglement distribution between users at different heights in a gravitational potential is affected by the curvature. Here, we numerically look at such effects on current and future quantum communication technologies.
It turns out that current regimes of operations for proposed satellite are based on technologies that are weakly affected by gravity. However, next-generation satellite missions may implement technologies that are based on narrowband optical systems which could experience substantial and measurable effects.

In this section we focus on regimes of operation in which the impact of space-time curvature on quantum communication protocols is significant.
Suitable candidates for a single photon source are cavity-enhanced spontaneous parametric down-conversion sources (cavity enhanced SPDCs) or atomic-vapor based single photon sources \cite{Kuklewicz:Wong:06,Jayakumar:Harishankar:13,Wolfgramm:deIcazaAstiz:11,Zhang:XianMin:11}. The regime of operation of interest, accessible by the current technology, is for center wavelengts of $\Omega_0=428$~nm or shorter and bandwidths of $\sigma=1$MHz or lower, where $\sigma\ll\Omega_0$. While this wavelength is significantly shorter than the the typical wavelengths for conventional optical sources, ranging from 780 nm (384 THz) to 1550 nm (193 THz) \cite{Buller:Collins:10}, it will be favorable for long distance free space transmission due to its low diffraction induced loss \cite{Bourgoin:Meyer-Scott:13}.

\subsection{Gaussian wave packets}
Let Alice and Bob employ single photon sources with such features. The normalized wave packets at both stations will have the form
\begin{eqnarray}
F_{\Omega_0}(\Omega)=\frac{1}{\sqrt[4]{2\pi\sigma^2}}e^{-\frac{(\Omega-\Omega_0)^2}{4\sigma^2}}\label{Bob:wave:packet},
\end{eqnarray}
where we have assumed the wave packet is real without loss of generality. As discussed in the previous section, the propagation of one photon in the gravitational field from Alice's station to Bob's station (or viceversa) will affect the shape of the photon's wave packet. In particular, if the photon was sent by Bob with wave packet $F^{(B)}_{\Omega_{B,0}}$ of the form \eqref{Bob:wave:packet}, it will be received by Alice as a photon with a wave packet $F^{(A)}_{\Omega_{A,0}}$ that differs to the original one and is related to $F^{(B)}_{\Omega_{B,0}}$ by \eqref{total:wave:packet:relation}.

We have shown that the mode overlap $\Delta$ quantifies the effects of gravity on the entanglement distribution, when photons are sent by one user and processed at a different location. In the case of the protocol considered in section \ref{protocol}, we need to compute $\Delta$ at Alice's station and express the Negativity $\mathcal{N}$ as a function of $\Delta$ through \eqref{scenario:negativity} and \eqref{overlap:fidelity:relation}.  We find
\begin{eqnarray}
\Delta=\int_{-\infty}^{+\infty}d\Omega_A\, F^{(A)}_{\Omega_{A,0}}(\Omega_A)F^{(B)}_{\Omega_{B,0}}(\Omega_A).
\end{eqnarray}
Note that the integral should be performed over stricly positive frequencies. However, since $\Omega_0\gg\sigma$, it is possible to include negative frequencies without affecting the value of $\Delta$. Using \eqref{Bob:wave:packet} and \eqref{total:wave:packet:relation}, simple algebra allows us to conclude that in our case
\begin{eqnarray}
\Delta=\sqrt{\frac{2(1\pm\delta)}{1+(1\pm\delta)^2}}e^{-\frac{\delta^2\Omega_{B,0}^2}{4(1+(1\pm\delta)^2)\sigma^2}}\label{final:result},
\end{eqnarray}
where we have defined
\begin{eqnarray}
\delta=\left|\sqrt[4]{\frac{1 - \frac{2M}{r_A}}{1 - \frac{3M}{r_B}}}-1\right|
\end{eqnarray}
and the signs $\pm$ occur for $r_B<r_A$ or $r_B>r_A$ respectively.
Notice that $\delta=0$ occurs either when Alice and Bob are in flat space-time ($f(r_A)=f(r_B)=1$) or Alice and Bob are at the same height ($f(r_A)=f(r_B)$). In both cases the modes perfectly  overlap ($\Delta=1$) as expected and there is no effect due to gravity.

Combining equation \eqref{final:result} with \eqref{scenario:negativity} we can predict how any protocol that depends explicitly on the mode overlap $\Delta$, for example the entanglement distribution protocol of section \ref{protocol}, is affected by the space-time channel.
We can use typical values for Earth to LEO communication and set $r_A=6371$km and $r_B=6771$km (i.e. the ISS orbit of about $400$km). Since the Schwarzschild radius of the Earth is $r_S=9$mm, we find that
\begin{eqnarray}
\delta\sim-\frac{1}{4}(\frac{r_s}{r_B}-\frac{r_s}{r_A})=1.45\times 10^{-11}.
\end{eqnarray}
We notice from \eqref{final:result} that two different scenarios can occur.
\begin{itemize}
	\item[i)] If $\frac{\delta\Omega_{B,0}}{\sigma}\leq\delta\ll1$ then
	\begin{eqnarray}
		\Delta\sim1-\mathcal{O}(\delta^2),
	\end{eqnarray}
	and therefore it is easy to see that $q\sim\delta^2\leq10^{-20}$. In this case the effects are independent of the peak frequency and on the width of the distribution and are negligible.
	\item[ii)] Surprisingly another scenario is possible, when $\delta\ll(\frac{\delta\Omega_{B,0}}{\sigma})^2\ll1$, which occurs for typical communication where $\Omega_{B,0}=700$THz (corresponding to a wavelength of about 420 nm) and $\sigma=1$MHz. For example, similar peak frequency and bandwidths have been achieved by trapped ion experiments \cite{Matsukevich:Maunz:08}. Then
	\begin{eqnarray}
		\Delta\sim1-\frac{\delta^2\Omega_{B,0}^2}{8\sigma^2}=1.3\times10^{-3},
	\end{eqnarray}
	and therefore $q\sim(\frac{\delta\Omega_{B,0}}{2\sigma})^2=2.6\times10^{-3}$. This effect is much larger than the one in the previous scenario and very close to the threshold of measurable effects with current technology.
	Furthermore, if Bob were to be very far from Earth ($f(r_B)=1$) then we would have $\delta=3.5\times10^{-10}$. With the same pulse characteristics we would achieve $q\sim(\frac{\delta\Omega_{B,0}}{2\sigma})^2=1.5\times10^{-2}$ which would be a $0.7$\% correction to the ideal Negativity $\mathcal{N}=1/2$ of flat space-time. This would produce measurable effect in the QBER of QKD protocols \cite{Nilsson:Stevenson:13}.
	
We can evaluate the effects of curvature on entanglement distribution for different types of sources. If Bob employed a Rb vapor type source with $\Omega_{B,0}\sim380$THz and $\sigma\sim5$MHz we find $q\sim 2.52\times 10^{-4}$, while for NV centres the effect is even smaller, $q\sim10^{-6}$.
\end{itemize}

\subsection{Impact on realistic communication protocols}
Two users Alice and Bob can employ QKD protocols to share a secret key. We assume that the users do not need to trust any node , source or device that is employed (device independent QKD). A relevant figure of merit for a QKD protocol is given by the QBER, defined as the ratio of exchanged error bits and the total number of sifted key bits. Using the entanglement distribution scheme of Fig. \ref{fig:SPS}, Alice and Bob can employ the QKD protocol proposed in \cite{Duan:Lukin:01} to share secret keys. Under the same assumptions as in section \ref{protocol:section}, in Appendix \ref{qber:appendix} we show that for such a protocol the $QBER\sim\frac{q}{2}$ which implies
\begin{eqnarray}
	\text{QBER}\sim\frac{\delta^2\Omega_{B,0}^2}{8\sigma^2}\label{qber}
\end{eqnarray}
which can reach the value $\sim0.7\%$ for $\Omega_0=480$nm and $\sigma=1$MHz.
This could be a noticeable effect in realistic implementations of QKD, which typically operate with QBERs of a few percent \cite{Nilsson:Stevenson:13}.

Equation \eqref{qber} accounts for effects due to only the curvature of space-time. The QBER that would be measured in realistic experiments must take into account other sources of errors, such as dark counts, channel losses, detector and sources imperfections, flatness across the spectrum of all devices (sources, beam splitters, detectors).

\section{Conclusions}
We have introduced mathematical techniques to study and quantify the effects of gravity on quantum information and quantum communication protocols. We have shown that  photon propagation is affected by the curvature of space-time, and may change their frequency distribution in centre, shape and bandwidth. We analyzed two different protocols, an entanglement distribution protocol and a continuous variable QKD protocol. We have shown that communications between two users that are located at different heights in the gravitational potential of the Earth are affected by the curvature of the space-time. Our results identify additional effects which cannot occur if two parties are situated at the same height or are in flat space-time. Therefore, the results of this paper unveil that there exist effects of gravity on quantum information protocols that cannot be reproduced and studied in Earth-based laboratories. These curvature effects would occur in addition to those due to special relativity or noise. 

While typical predictions of quantum field theory, such as the Dynamical Casimir effect, require enormous accelerations \cite{Wilson:Johansson:11}, the effects studied in this paper are therefore relevant for possible space based implementations of satellite missions based on current technologies and could potentially be tested within near-future proposals for satellite missions. While these effects could in principle be compensated by exchange of additional resources (use of local oscillators, tunable receiving devices, tunable sources), they will help us to investigate the overlap between quantum mechanics and the theory of general relativity.

\section*{Acknowledgements}
We thank Daniel K. Oi, Jason Doukas, Mehdi Ahmadi, Nicolai Friis, Antony Lee, Jorma Louko and Paolo Villoresi for useful discussions and comments.
This work was in part supported by the UK Engineering and Physical Science Research Council grant number EP/J005762/1 and the European Community's Seventh Framework Programme under Grant Agreement 277110. T. Jennewein acknowledges support from NSERC, CIFAR, CFI, the Ontario Ministry for  Science, Industry Canada, CSA. D. Bruschi would like to thank the university of Nottingham for hospitality. I. Fuentes acknowledges support from EPSRC (CAF Grant No. EP/G00496X/2).

\appendix

\section{QBER\label{qber:appendix}}
We apply our results to a well known scheme such as the one described in \cite{Duan:Lukin:01}. There, Alice and Bob have two memories each, $A,A^{\prime}$ and $B,B^{\prime}$ respectively. They use the scheme in Section \ref{curved:spacetime:protocol:section} to entangle $A$ with $B$ and $A^{\prime}$ with $B^{\prime}$. The modes stored in the memories $A,A^{\prime}$ can then be mixed at a balanced beamsplitter, and analogously for the modes in memories $B,B^{\prime}$. If one detector per user clicks the distribution protocol has been successful. This occurs on average on $50$\% of the cases. Alice and Bob assign a bit value to each detector. Alice and Bob share the same bit if the state of the memories $A,B$ and the state of the memories $A^{\prime},B^{\prime}$ are the same (i.e., both $\rho_+$ or $\rho_-$). This occurs on average with probability $p_{\text{share}}=\frac{(1-\sqrt{1-q})^2}{4}+\frac{(1+\sqrt{1-q})^2}{4}$. The probability of Alice and Bob not sharing the same bit is instead $p_{\text{diff}}=2\frac{(1+\sqrt{1-q})}{2}\frac{(1-\sqrt{1-q})}{2}$.

The QBER is defined as the number of different bits shared by Alice and Bob over the total bits exchanged, namely QBER$:=\frac{p_{\text{diff}}}{p_{\text{share}}+p_{\text{diff}}}$. Substituting for $p_{\text{share}}$ and $p_{\text{diff}}$ we find
\begin{eqnarray*}
	\text{QBER}=\frac{q}{2}.
\end{eqnarray*}

\bibliography{Biblio}
%\bibliographystyle{unsrt}

%\begin{thebibliography}{99}
%\end{thebibliography}

\end{document}